\documentclass[5p,authoryear,sort&compress]{elsarticle}
\usepackage{mathpazo}
\usepackage{booktabs}
\usepackage{multirow}
\usepackage{color}
\usepackage{url}

\usepackage{amssymb}

\usepackage[switch]{lineno}

\journal{Atmospheric Research}

\begin{document}

\begin{frontmatter}



\title{Origin of atmospheric aerosols at the Pierre Auger Observatory using studies of air mass trajectories in South America}


\author[label1]{The Pierre Auger Collaboration}
\author[label2]{Gabriele Curci}

\address[label1]{Observatorio Pierre Auger, Av. San Mart\'in Norte 304, 5613 Malarg\"ue, Argentina}
\address[label2]{CETEMPS, Department of Physics, University of L'Aquila, L'Aquila, Italy}
\begin{abstract}
The Pierre Auger Observatory is making significant contributions towards understanding the nature and origin of ultra-high energy cosmic rays. One of its main challenges is the monitoring of the atmosphere, both in terms of its state variables and its optical properties. The aim of this work is to analyze aerosol optical depth $\tau_{\rm a}(z)$ values measured from 2004 to 2012 at the observatory, which is located in a remote and relatively unstudied area of the Pampa Amarilla, Argentina. The aerosol optical depth is in average quite low -- annual mean $\tau_{\rm a}(3.5~{\rm km})\sim 0.04$ -- and shows a seasonal trend with a winter minimum -- $\tau_{\rm a}(3.5~{\rm km})\sim 0.03$ --, and a summer maximum -- $\tau_{\rm a}(3.5~{\rm km})\sim 0.06$ --, and an unexpected increase from August to September -- $\tau_{\rm a}(3.5~{\rm km})\sim 0.055$). We computed backward trajectories for the years 2005 to 2012 to interpret the air mass origin. Winter nights with low aerosol concentrations show air masses originating from the Pacific Ocean. Average concentrations are affected by continental sources (wind-blown dust and urban pollution), while the peak observed in September and October could be linked to biomass burning in the northern part of Argentina or air pollution coming from surrounding urban areas.
\end{abstract}


\end{frontmatter}


\section{Introduction}
Modelling of aerosols in climate models is still a challenging task, also due to the lack of a complete global coverage of long-term ground-based measurements. In South America, only few studies have been done, usually located in mega-cities~\citep{SA_1,SA_2,SA_3,SA_4,SA_5}. Astrophysical observatories need a continuous monitoring of the atmosphere, including aerosols, and thus offer an unique opportunity to get a characterisation of aerosols in the same locations over several years. Here we report on seven years of aerosol optical depth measurements carried out at the Pierre Auger Observatory in Argentina.

The Pierre Auger Observatory is the largest operating cosmic ray observatory ever built~\citep{PAO_1,PAO_2}. It is conceived to measure the flux, arrival direction distribution and mass composition of cosmic rays from $10^{18}~$eV to the very highest energies. It is located in the Pampa Amarilla ($35.1^\circ - 35.5^\circ~${S}, $69.0^\circ - 69.6^\circ~${W}, and $1\,300-1\,700~$m above sea level), close to Malarg\"{u}e, Province of Mendoza. Construction was completed at the end of 2008 and data taking for the growing detector array started at the beginning of 2004. The observatory consists of about $1660$~surface stations -- water-Cherenkov tanks and their associated electronics -- covering an area of $3000~$km$^2$. In addition, $27~$telescopes, housed in four fluorescence detector (FD) buildings, detect air-fluorescence light above the array during nights with low-illuminated moon and clear optical conditions. The atmosphere is used as a giant calorimeter, representing a detector volume larger than $30\,000~$km$^3$. Once cosmic rays enter into the atmosphere, they induce extensive air showers of secondary particles. Charged particles of the shower excite atmospheric nitrogen molecules, and these molecules then emit fluorescence light mainly in the $300-420~$nm wavelength range. The number of fluorescence photons produced is proportional to the energy deposited in the atmosphere through electromagnetic energy losses undergone by the charged particles. Then, from their production point to the telescope, photons can be scattered by molecules ({\it Rayleigh scattering}) and/or atmospheric aerosols ({\it Mie scattering}). A small component (at shorter ultra-violet wavelengths) of the fluorescence light can be absorbed by some atmospheric gases such as ozone or nitrogen dioxide. 

\begin{figure}[!t]
\centering
	\includegraphics [scale=0.40]{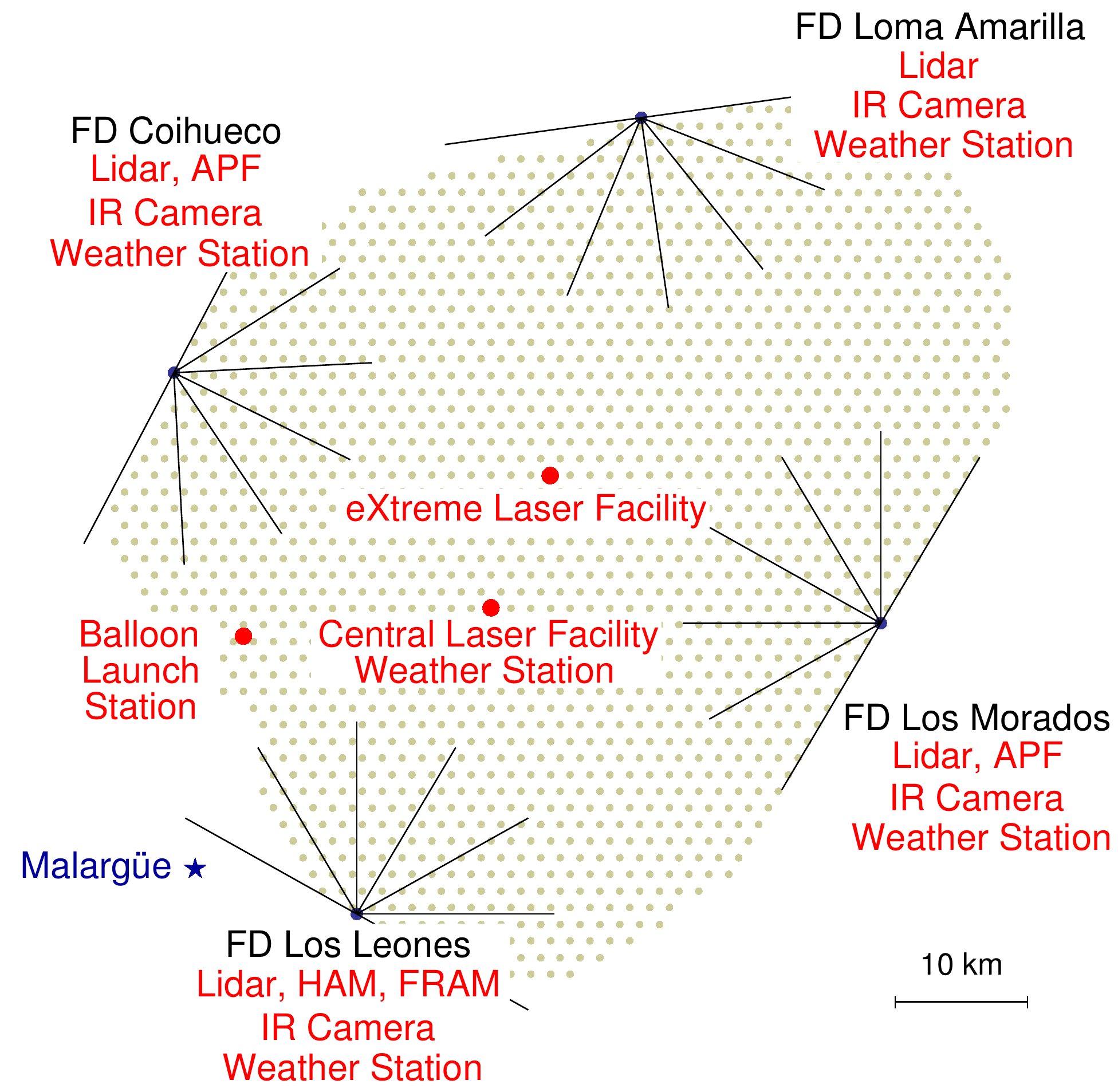}
\caption {{\bf Atmospheric monitoring map of the Pierre Auger Observatory (from~\citet{AugerATMO_LongPaper}).} Gray dots indicate the positions of surface detector (SD) stations. Black segments indicate the fields of view of the fluorescence detectors (FD) which are located in four sites, called Los Leones, Los Morados, Loma Amarilla and Coihueco, on the perimeter of the surface array. Each FD site hosts several atmospheric monitoring facilities.}
\label{fig:AugerArray}
\end{figure}

The aerosol component is the most variable term contributing to the atmospheric transmission function. Thus, to reduce as much as possible the systematic uncertainties on air shower reconstruction using the fluorescence technique, aerosols have to be continuously monitored. An extensive atmospheric monitoring system has been developed at the Pierre Auger Observatory~\citep{AugerATMO_LongPaper,MyEPJP}. The different facilities and their locations are shown in Figure~\ref{fig:AugerArray}. Aerosol monitoring is performed using two central lasers (CLF / XLF)~\citep{CLF_jinst}, four elastic scattering lidar stations~\citep{Lidar_BenZvi}, two aerosol phase function monitors (APF)~\citep{APF_BenZvi} and two setups for the \r{A}ngstr\"om parameter, the Horizontal Attenuation Monitor (HAM)~\citep{HAM_proceeding} and the Photometric Robotic Atmospheric Monitor (FRAM) \citep{FRAM_proceeding}. Also, a Raman lidar is operational {\it in-situ} since June 2013. In Section~\ref{sec:vaod_measurements}, the measurements of the aerosol optical depth are described. The HYSPLIT air-modelling programme will be briefly described in Section~\ref{sec:backtrajectory}, together with a detailed view on the air mass trajectories and origin of the aerosols passing above the Pierre Auger Observatory. Finally, aerosol measurements and their different features will be interpreted using backward trajectories of air masses in Section~\ref{sec:interpret_aerosol_hysplit}. A preliminary version of this work was presented in~\citet{MyECRS}, showing some links between air mass trajectories and aerosol measurements at the Pierre Auger Observatory. This paper provides a more complete study with the full data set available for aerosol measurements.

\begin{figure*}[!t]
\centering
	\includegraphics [scale=1.0]{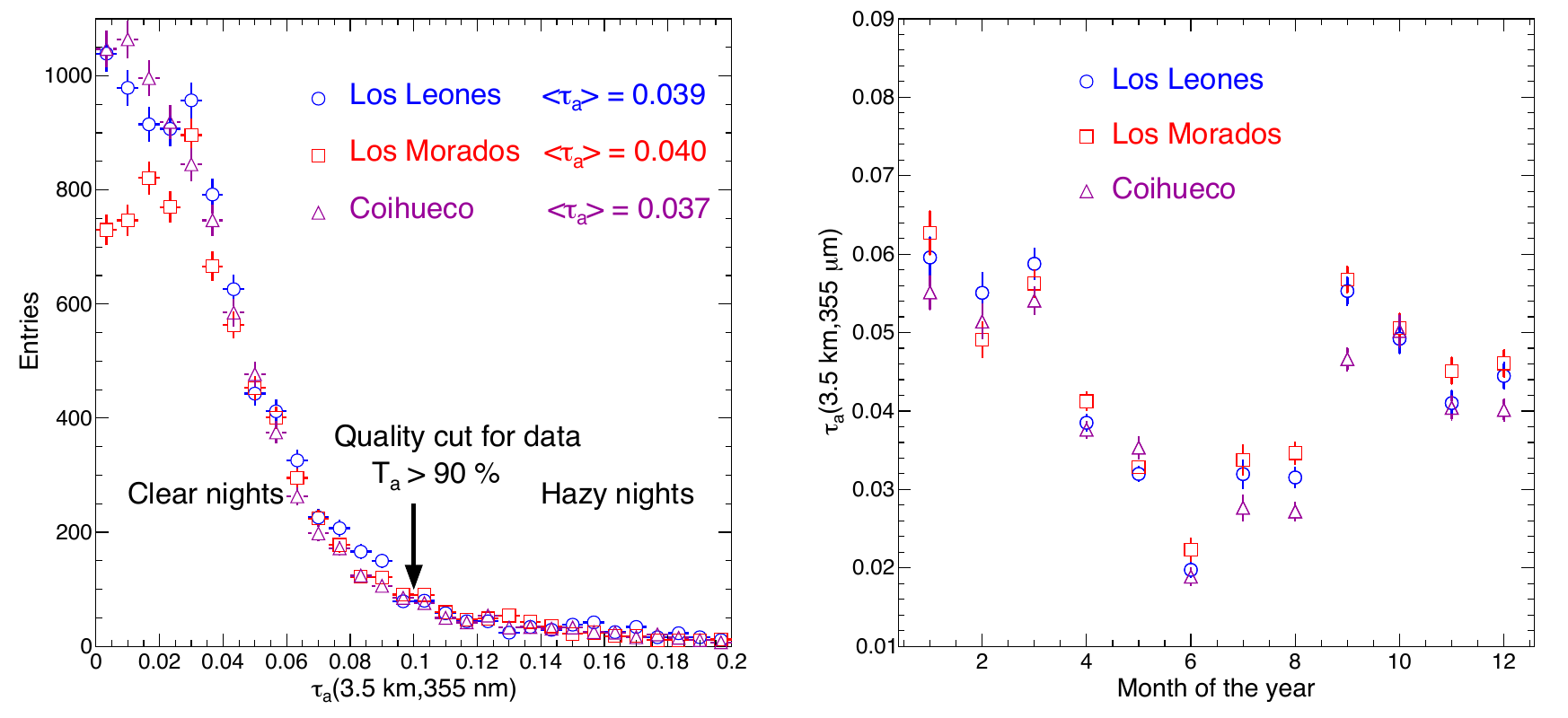}
\caption{{\bf Vertical aerosol optical depth measurements integrated from the ground up to $3.5~$km at the fluorescence telescopes Los Leones, Los Morados and Coihueco.} (left) Distribution of aerosol optical depth values. (right) Monthly variation of mean aerosol optical depth values. Data set between January 2004 and December 2012 in both figures. The horizontal error bars correspond to the bin size, and the vertical error bars represent the root mean square of the mean value. Colours in online version.}
\label{fig:CLF_VAOD}
\end{figure*}

\section{Aerosol optical depth measurements}
\label{sec:vaod_measurements}
At the Pierre Auger Observatory, several facilities have been installed to monitor the aerosol component of the atmosphere. One of the aerosol measurements made at the observatory is the aerosol optical depth using laser tracks generated by the Central Laser Facility. This facility is operated only at nights when the observatory is taking data: thus, aerosol data obtained are more a sampling data set than continuous measurements. The CLF is located in a position equidistant from three out of four FD sites. The main component is a laser producing a beam with a wavelength $\lambda_0$ fixed at $355~$nm, i.e.\ in the middle of the nitrogen fluorescence spectrum emitted by nitrogen molecules excited by the passing of air showers. The pulse width of the beam is $7$~ns and a maximum energy per pulse is of $7$~mJ. This corresponds to the fluorescence light produced by an air shower with an energy of $10^{20}~$eV viewed from a distance of $20~$km. Typically, the beam is directed vertical. When a laser shot is fired, the fluorescence telescope detects a small fraction of the light scattered out of the laser beam. The recorded signal depends on the atmospheric properties. Two methods have been developed by the Auger Collaboration to estimate hourly the vertical aerosol optical depth $\tau_{\rm a}(h,\lambda_0)$ with the CLF, where $h$ is the altitude above ground level and $\lambda_0$ the CLF wavelength. Both methods assume a horizontal uniformity for the molecular and aerosol components. The first method, the so-called "Data Normalised Analysis", is an iterative procedure comparing hourly average light profiles to a reference clear night where light attenuation is dominated by molecular scattering. Using a reference clear night avoids an absolute photometric calibration of the laser. About one reference clear night per year per FD site is found to be sufficient. The second method, the so-called "Laser Simulation Analysis", is based on the comparison of measured laser light profiles to profiles simulated with different aerosol attenuation conditions defined using a two-parameters model. More details can be found in \citet{CLF_valore}.

\begin{table*}[!t]
\centering
\begin{tabular}{c | c c c c c c c c c c c c}
\toprule
Year & Jan & Feb & Mar & Apr & May & Jun & Jul & Aug & Sep & Oct & Nov & Dec \\
\midrule
\multicolumn{13}{c} {\bf Clear hourly profiles} \\[0.5ex]
\midrule
\multirow{2}{*}{2005} & -- & -- & -- & 2\% & 20\% & {\color{blue}58\%} & {\color{blue}38\%} & 5\% & 21\% & 4\% & 4\% & 11\%  \\
& -- & -- & -- & 1 & 8 & {\color{blue}7} & {\color{blue}16} & 1 & 5 & 3 & 2 & 4  \\[0.5ex]
\multirow{2}{*}{2006} & 0\% & 0\% & 0\% & 2\% & 4\% & {\color{blue}29\%} & {\color{blue}29\%} & 4\% & 14\% & 14\% & 17\% & 9\% \\
& 0 & 0 & 0 & 2 & 4 & {\color{blue}33} & {\color{blue}31} & 6 & 17 & 3 & 14 & 7 \\[0.5ex]
\multirow{2}{*}{2007} & 0\% & 0\% & 3\% & 31\% & 10\% & {\color{blue}54\%} & {\color{blue}88\%} & {\color{blue}63\%} & 7\% & 27\% & 36\% & 14\% \\
& 0 & 0 & 2 & 30 & 10 & {\color{blue}42} & {\color{blue}35} & {\color{blue}42} & 5 & 25 & 36 & 10 \\[0.5ex]
\multirow{2}{*}{2008} & 20\% & 11\% & 3\% & 10\% & 1\% & {\color{blue}53\%} & 14\% & {\color{blue}39\%} & 10\% & 5\% & 7\% & 0\% \\
& 11 & 7 & 2 & 9 & 1 & {\color{blue}53} & 16 & {\color{blue}32} & 9 & 5 & 5 & 0 \\[0.5ex]
\multirow{2}{*}{2009} & 4\% & 2\% & 0\% & 0\% & 21\% & {\color{blue}53\%} & 20\% & 11\% & 4\% & 22\% & 17\% & 3\% \\
& 3 & 2 & 0 & 0 & 12 & {\color{blue}57} & 17 & 10 & 4 & 20 & 15 & 2 \\[0.5ex]
\multirow{2}{*}{2010} & 1\% & 15\% & 5\% & 14\% & 13\% & {\color{blue}41\%} & {\color{blue}35\%} & 4\% & 0\% & 1\% & 0\% & 8\%\\
& 1 & 9 & 5 & 20 & 14 & {\color{blue}34} & {\color{blue}33} & 4 & 0 & 1 & 0 & 5 \\[0.5ex]
\multirow{2}{*}{2011} & 5\% & 0\% & 5\% & 8\% & 14\% & {\color{blue}32\%} & {\color{blue}48\%} & {\color{blue}27\%} & 3\% & 0\% & 7\% & 8\%\\
& 3 & 0 & 7 & 11 & 17 & {\color{blue}34} & {\color{blue}30} & {\color{blue}28} & 3 & 0 & 6 & 6 \\[0.5ex]
\multirow{2}{*}{2012} & 5\% & 13\% & 2\% & 3\% & 16\% & {\color{blue}57\%} & {\color{blue}33\%} & 15\% & 0\% & 8\% & 0\% & 7\%\\
& 4 & 9 & 2 & 3 & 21 & {\color{blue}59} & {\color{blue}35} & 17 & 0 & 7 & 0 & 5 \\[0.5ex]
\multirow{2}{*}{All} & 5\% & 6\% & 3\% & 9\% & 11\% & {\color{blue}45\%} & {\color{blue}33\%} & {\color{blue}20\%} & 6\% & 10\% & 13\% & 8\% \\
& 22 & 27 & 18 & 76 & 87 & {\color{blue}319} & {\color{blue}213} & {\color{blue}140} & 43 & 64 & 78 & 39 \\[0.5ex]
\midrule
\multicolumn{13}{c} {\bf Hazy hourly profiles} \\[0.5ex]
\midrule
\multirow{2}{*}{2005} & -- & -- & -- & 0\% & 0\% & 0\% & 0\% & 0\% & 4\% & 0\% & 6\% & 14\%  \\
& -- & -- & -- & 0 & 0 & 0 & 0 & 0 & 1 & 0 & 3 & 5 \\[0.5ex]
\multirow{2}{*}{2006} & {\color{red}40\%} & 0\% & 1\% & {\color{red}16\%} & 6\% & 3\% & 1\% & 6\% & 6\% & 0\% & {\color{red}11\%} & 0\% \\
& {\color{red}2} & 0 & 1 & {\color{red}18} & 6 & 3 & 1 & 8 & 7 & 0 & {\color{red}9} & 0  \\[0.5ex]
\multirow{2}{*}{2007} & {\color{red}16\%} & 6\% & 8\% & 6\% & 0\% & 1\% & 0\% & 1\% & {\color{red}10\%} & {\color{red}9\%} & 0\% & 7\% \\
& {\color{red}11} & 3 & 5 & 6 & 0 & 1 & 0 & 1 & {\color{red}7} & {\color{red}8} & 0 & 5 \\[0.5ex]
\multirow{2}{*}{2008} & 2\% & 8\% & 13\% & 1\% & 0\% & 0\% & {\color{red}26\%} & 1\% & {\color{red}30\%} & 1\% & 7\% & 18\%  \\
& 1 & 5 & 8 & 1 & 0 & 0 & {\color{red}30} & 1 & {\color{red}28} & 1 & 5 & 5  \\[0.5ex]
\multirow{2}{*}{2009} & {\color{red}38\%} & 9\% & {\color{red}21\%} & 2\% & 0\% & 4\% & {\color{red}20\%} & 5\% & 9\% & 0\% & 3\% & 12\%  \\
& {\color{red}26} & 9 & {\color{red}25} & 3 & 0 & 4 & {\color{red}17} & 5 & 9 & 0 & 3 & 9\\[0.5ex]
\multirow{2}{*}{2010} & 6\% & 6\% & 5\% & 5\% & 0\% & 1\% & 4\% & 3\% & 7\% & {\color{red}29\%} & 0\% & 0\% \\
& 5 & 4 & 5 & 7 & 0 & 1 & 4 & 3 & 8 & {\color{red}30} & 0 & 0 \\[0.5ex]
\multirow{2}{*}{2011} & 9\% & 4\% & 14\% & 2\% & 1\% & 2\% & 0\% & 3\% & {\color{red}24\%} & {\color{red}49\%} & {\color{red}21\%} & 14\% \\
& 6 & 4 & 19 & 3 & 1 & 2 & 0 & 3 & {\color{red}23} & {\color{red}34} & {\color{red}19} & 10 \\[0.5ex]
\multirow{2}{*}{2012} & 11\% & 10\% & 11\% & {\color{red}14\%} & 1\% & 0\% & 4\% & 6\% & {\color{red}15\%} & 6\% & {\color{red}21\%} & 0\% \\
& 8 & 7 & 11 & {\color{red}14} & 1 & 0 & 4 & 7 & {\color{red}14} & 5 & {\color{red}15} & 0 \\[0.5ex]
\multirow{2}{*}{All} & {\color{red}14\%} & 7\% & 11\% & 6\% & 1\% & 2\% & 9\% & 4\% & {\color{red}14\%} & 12\% & 9\% & 7\%  \\
& {\color{red}59} & 32 & 74 & 52 & 8 & 11 & 56 & 28 & {\color{red}97} & 78 & 54 & 34 \\[1.0ex]
\bottomrule
\end{tabular}
~\\
\caption{\label{tab:vaod_stats}{\bf Fraction and statistics of aerosol hourly profiles for clear and hazy aerosol conditions for each month between 2005 and 2012.} For each year, the first line gives the fraction of profiles corresponding to the aerosol conditions in the whole set of profiles recorded during the corresponding month. The second line gives the number of profiles associated to their corresponding fraction. Months without data are indicated by ``--". For each year, the two or three months with the highest fraction of clear and hazy nights are coloured in blue and red, respectively. Colours in online version.}
\end{table*}

\begin{figure}[!t]
\centering
\vspace{-0.8cm}
	\includegraphics [width=0.49\textwidth]{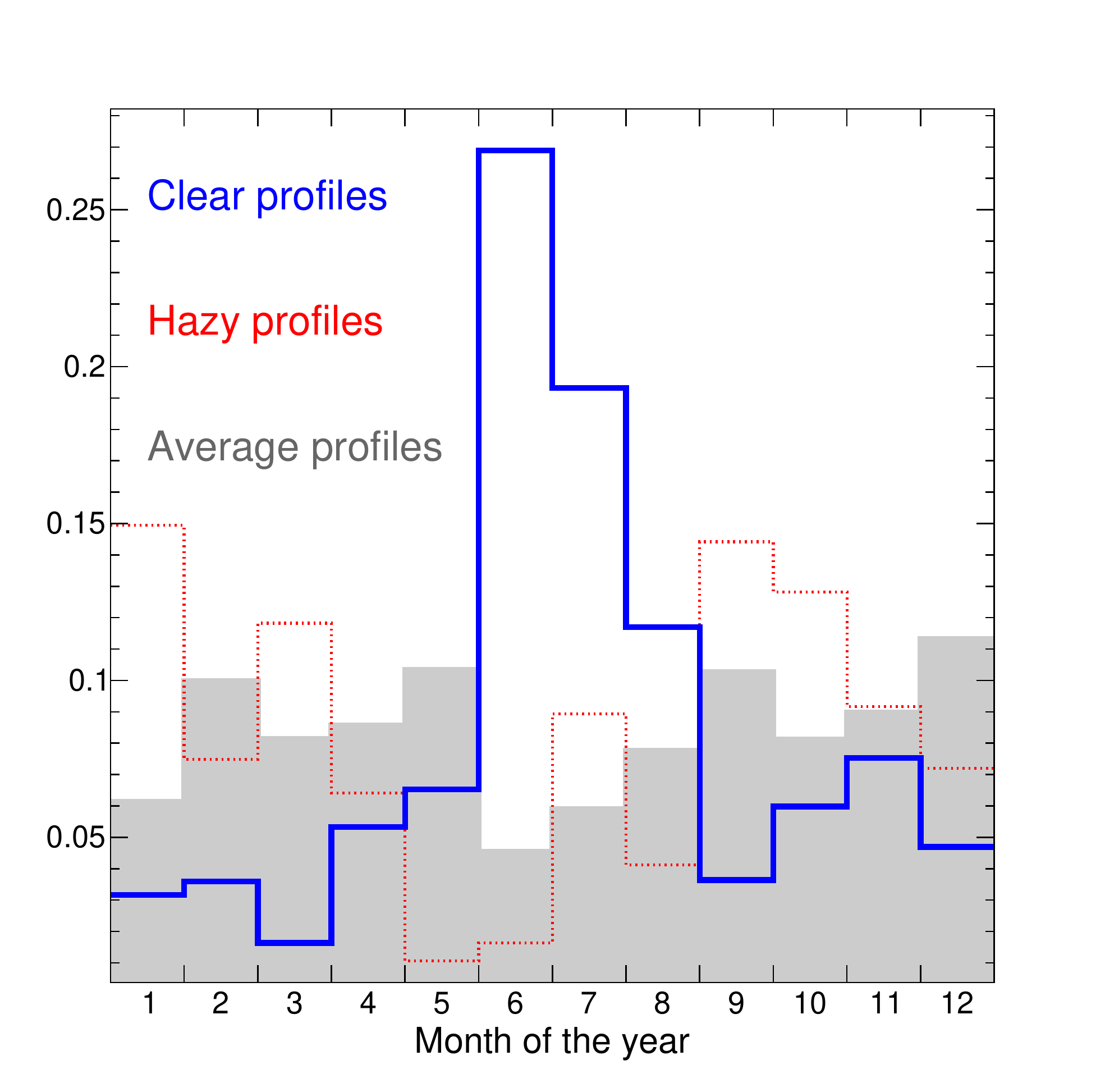}
\caption{{\bf Monthly frequency over a year of clear hourly aerosol profiles ($\tau_{\rm a}(3.5~{\rm km})\leqslant0.01$, {\it solid line}), average hourly aerosol profiles ($0.03< \tau_{\rm a}(3.5~{\rm km})\leqslant0.05$, {\it grey filled}) and hazy hourly aerosol profiles ($\tau_{\rm a}(3.5~{\rm km})>0.10$, {\it dotted line}) at Los Morados.} Data set between January 2005 and December 2012 is used here. Each bin is re-weighted to take into account the fact that not the same number of aerosol profiles is recorded in Winter (longer nights) or during Summer (shorter nights). Colours in online version.}
\label{fig:VAOD_values}
\end{figure}

The CLF provides hourly altitude profiles for each fluorescence site during fluorescence data acquisition. In Figure~\ref{fig:CLF_VAOD}~(left), the distribution of the aerosol optical depth integrated from the ground up to $3.5~$km above ground level, recorded at Los Leones, Los Morados and Coihueco is shown. Due to large distance to the CLF site, measurements from Loma Amarilla have not been included in this study. Only recently, data from the closer XLF have been used to measure the aerosol attenuation from Loma Amarilla~\citep{Laura_ICRC}. The mean value of $\tau_{\rm a}(3.5~{\rm km})$ is about 0.04. Nights with $\tau_{\rm a}(3.5$ ${\rm km})$ larger than $0.1$, meaning a transmission factor lower than $90\%$, are rejected for air shower studies at the Pierre Auger Observatory. Systematic uncertainties associated with the measurement of the aerosol optical depth are due to the relative calibration of the telescopes and the central laser, and the relative uncertainty of the determination of the reference clear profile. The total uncertainty is estimated to 0.006 for an altitude of 3.5~km above ground level. Figure~\ref{fig:CLF_VAOD}~(right) displays the monthly variation of the aerosol optical depth integrated up to $3.5~$km above ground level. The aerosol concentration depicts a seasonal trend, reaching a minimum during Austral winter and a maximum in Austral summer. This trend is typical and has already been observed in many long-term aerosol analyses~\citep{CastroVidelaEtAl,Benavente,SchaferEtAl}. Since spatial resolution in latitude and longitude for atmospheric data used by the HYSPLIT programme is one degree, only aerosol data measured at Los Morados are used since it is the closest fluorescence site to the coordinates ($35^\circ$ {{S}}, $69^\circ$ {{W}}). We subdivide the data sample into three populations:
\begin{itemize}
	\item {\bf the clear hourly aerosol profiles} with the lowest aerosol concentrations: $\tau_{\rm a}(3.5~{\rm km})\leqslant0.01$ ({\it 1126~trajectories}),
	\item {\bf the hazy hourly aerosol profiles} with the highest aerosol concentrations: $\tau_{\rm a}(3.5~{\rm km})>0.10$ ({\it 583~trajectories}),
	\item {\bf the average hourly aerosol profiles} with the average aerosol concentrations: $0.03< \tau_{\rm a}(3.5~{\rm km})\leqslant0.05$ ({\it 1918~trajectories}).
\end{itemize}
The relative frequencies month-by-month for clear conditions and hazy conditions are shown in Figure~\ref{fig:VAOD_values}. Clear conditions are more common during the Austral winter than in the rest of the year. Furthermore, a clear increase for the population of hazy aerosol profiles from August to September can be seen in both Figures~\ref{fig:CLF_VAOD}~(right) and~\ref{fig:VAOD_values}, contrary to the overall seasonal trend. Table~\ref{tab:vaod_stats} lists the fraction of clear and hazy aerosol profiles for each year between 2005 and 2012. For each year, the two or three highest fraction values are indicated. Clear conditions are very common during Austral winter throughout all years of this analysis. The unexpected peak in hazy conditions during September and October is almost as stable as the seasonal trend throughout the years, but not with the same statistical significance. It could be a consequence of biomass burning in the northern part of South America or closer pollution sources coming from the larger cities San Rafael and Mendoza in the North.

\section{Backward trajectory of air masses}
\label{sec:backtrajectory}
After having briefly presented the air-modelling programme used in this study and checked the validity of its calculations using meteorological radio soundings, this section aims for characterising the air masses crossing over the Pierre Auger Observatory.

\subsection{HYSPLIT -- an air-modelling programme}
\label{sec:hysplit}
Different models have been developed to study air mass relationships between two regions. Among them, the {\it HYbrid Single-Particle Lagrangian Integrated Trajectory} model, or HYSPLIT~\citep{HYSPLIT_1, HYSPLIT_2}, is a commonly used air-modelling programme in atmospheric sciences for calculating air mass displacements from one region to another. The HYSPLIT model, developed by the Air Resources Laboratory, NOAA\footnote{NOAA, National Oceanic and Atmospheric Administration, U.S.A.}, is a complete system designed to support a wide range of simulations related to regional or long-range transport and dispersion of airborne particles~\citep{MOCAGE}. It is possible to compute simple trajectories for complex dispersion and deposition simulations using either puff or particle approaches within a Lagrangian framework~\citep{DeVito,Cao}. In this work, HYSPLIT will be used to get backward / forward trajectories by tracking air masses backward / forward in time. The resulting backward / forward trajectory indicates air mass arriving at a specific time in a specific geographical location (latitude, longitude and altitude), identifying the regions linked to it. All along the air mass paths, hourly meteorological data are used. Trajectory uncertainty for computed air masses is usually divided into three components: {\it the physical uncertainty} due to the inadequacy of the representation of the atmosphere in space and time by the model, {\it the computational uncertainty} due to numerical uncertainties and {\it the measurement uncertainty} for the meteorological data field in creating the model. Also, there could be sensitivity to initial conditions, especially during periods with large instabilities: for instance, estimation of the beginning of backward trajectories could be affected by very turbulent and chaotic air mass movement.

HYSPLIT provides details of some of the meteorological parameters along the trajectory. It is possible to extract information on terrain height, pressure, ambient and potential temperature, rainfall, relative humidity and solar radiation. However, to produce a trajectory, HYSPLIT requires at least the wind vector, ambient temperature, surface pressure and height data. These data can come from different meteorological models. Among the available models in HYSPLIT, the most used are the North American Meso (NAM), the NAM Data Assimilation System (EDAS) and the Global Data Assimilation System (GDAS). Only the GDAS model provides meteorological data for the site of the Pierre Auger Observatory for the period starting in January 2005 and extending to the present time. The GDAS is an atmospheric model developed by the NOAA~\citep{GDAS}. Those data are distributed over a one degree latitude/longitude grid ($360^\circ \times 180^\circ$), with a temporal resolution of three hours. GDAS provides 23 pressure levels, from $1000~$hPa (more or less sea level) to $20~$hPa (about 26~km altitude). The data set is complemented by data for the surface level at the given location. The GDAS grid point most suitable for the location of the Pierre Auger Observatory is ($35^\circ$ {{S}}, $69^\circ$ {{W}}), i.e.\ just slightly inside the array to the north-east~\citep{GDASpaper}. Lateral homogeneity of the atmospheric variables across the Auger array is assumed~\citep{AugerATMO_LongPaper}. Validity of GDAS data was previously studied by the Auger Collaboration: the agreement with ground-based weather station and meteorological radiosonde launches has been verified. The work consisted of comparing the temperature, humidity and pressure values with those measured by the monitoring systems at Auger. For instance, distributions of the differences between measured weather station data at the centre of the array and GDAS model data were obtained for temperature, pressure and water vapour pressure: $1.3~$K $[{\sigma}=3.9~{\rm K}]$, $0.4~$hPa $[{\sigma}=1.2~{\rm hPa}]$ and $-0.2~$hPa $[{\sigma}=2.1~{\rm hPa}]$, respectively~\citep{GDASpaper}. Thanks to their highly reliable availability and high frequency of data sets, it was concluded that GDAS data could be employed as a suitable replacement for local weather data in air shower analyses of the Pierre Auger Observatory. The agreement between GDAS model and local measurements has been checked only for state variables of the atmosphere. In the next section, wind data, a key parameter in the HYSPLIT model, are tested.

\subsection{Validity of the HYSPLIT calculations using meteorological radio soundings}
\label{sec:validity_balloon}

\begin{figure*}
\centering
	\includegraphics [scale=1.0]{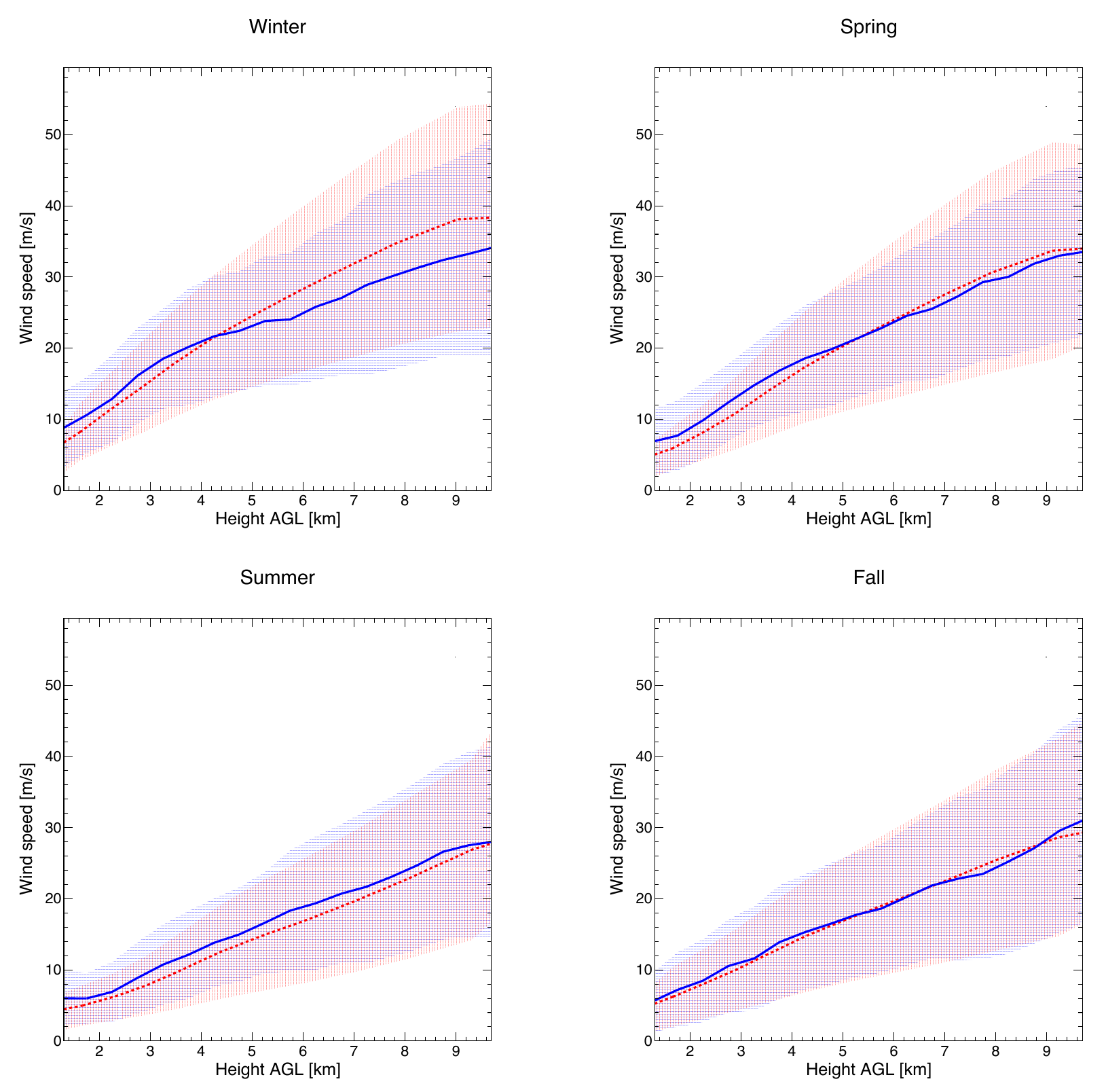}
\caption{{\bf Seasonal vertical profiles for wind speed using the radio soundings and the GDAS model at the Malarg\"ue location.} Radio soundings data {\it (solid line)} from August 2002 to December 2010 are used: 72 profiles in winter, 95 profiles in spring, 81 profiles in summer and 81 profiles in fall. GDAS data {\it (dashed line)} from January 2005 to December 2010 are used. Each seasonal profile contains approximately 4000~profiles. The curves represent the averaged profile of wind speed for the corresponding season. The dashed bands show the distribution of 68\% of the measurements. Colours in online version.}
\label{fig:Wind_profile_seasonal}
\end{figure*}


\begin{table*}
\centering
\begin{tabular}{c | c c c c | c c c c}
\toprule
&\multicolumn{4}{c} {\bf $\langle X_{\rm GDAS} - X_{\rm RS} \rangle$} & \multicolumn{4}{c} {\bf ${\rm RMS}(X_{\rm GDAS} - X_{\rm RS})$} \\[0.5ex]
\midrule
Altitude AGL [m] & $T$ [K] & $p$ [hPa] & $e$ [hPa] & $v_{\rm w}$ [m/s] & $T$ [K] & $p$ [hPa] & $e$ [hPa] & $v_{\rm w}$ [m/s] \\[0.5ex]
\midrule
1000  & $-0.82$ & $-0.35$ & $-0.23$ & $-1.00$ & $0.20$ & $0.10$ & $0.18$ & $1.55$ \\
2000  & $-0.10$ & $-0.50$ & $-0.05$ & $-0.50$ & $0.15$ & $0.05$ & $0.09$ & $1.20$ \\
4000  & $+0.50$ & $-0.20$ & $+0.02$ & $-1.30$ & $0.10$ & $0.03$ & $0.02$ & $1.75$ \\
6000  & $+0.08$ & $-0.02$ & $+0.05$ & $+1.40$ & $0.07$ & $0.02$ & $0.01$ & $2.84$ \\
8000  & $+0.20$ & $+0.01$ & $+0.02$ & $+2.50$ & $0.08$ & $0.02$ & $0.02$ & $3.45$ \\ [1.0ex]
\bottomrule
\end{tabular}
~\\
\caption{\label {tab:stat_agree}{\bf Mean and standard deviation values for the differences between measured radio sounding (RS) data and GDAS model data.} Values calculated for different probed altitudes. $T$: temperature, $p$: pressure, $e$: vapour pressure, $v_{\rm w}$: wind speed. Data sets from January 2005 to December 2010.}
\end{table*}

Above the Pierre Auger Observatory, the height dependent profiles have been measured using meteorological radiosondes launched mainly from the Balloon Launch Station (BLS, Figure~\ref{fig:AugerArray}). The balloon flight programme was terminated in December 2010 after having been operated 331 times since August 2002~\citep{RapidMonitoring,EPJP_GDAS}. The radiosonde records data every $20~$m, approximately, up to an average altitude of $25~$km above sea level, well above the fiducial volume of the fluorescence detector. The average time elapsed during its ascent was about $100~$minutes on average. The measurement accuracies are $0.2^\circ$C in temperature, $0.5-1.0~$hPa in pressure and $5\%$ in relative humidity.

\begin{figure}[!t]
\centering
\vspace{-0.3cm}
	\includegraphics [scale=1.0]{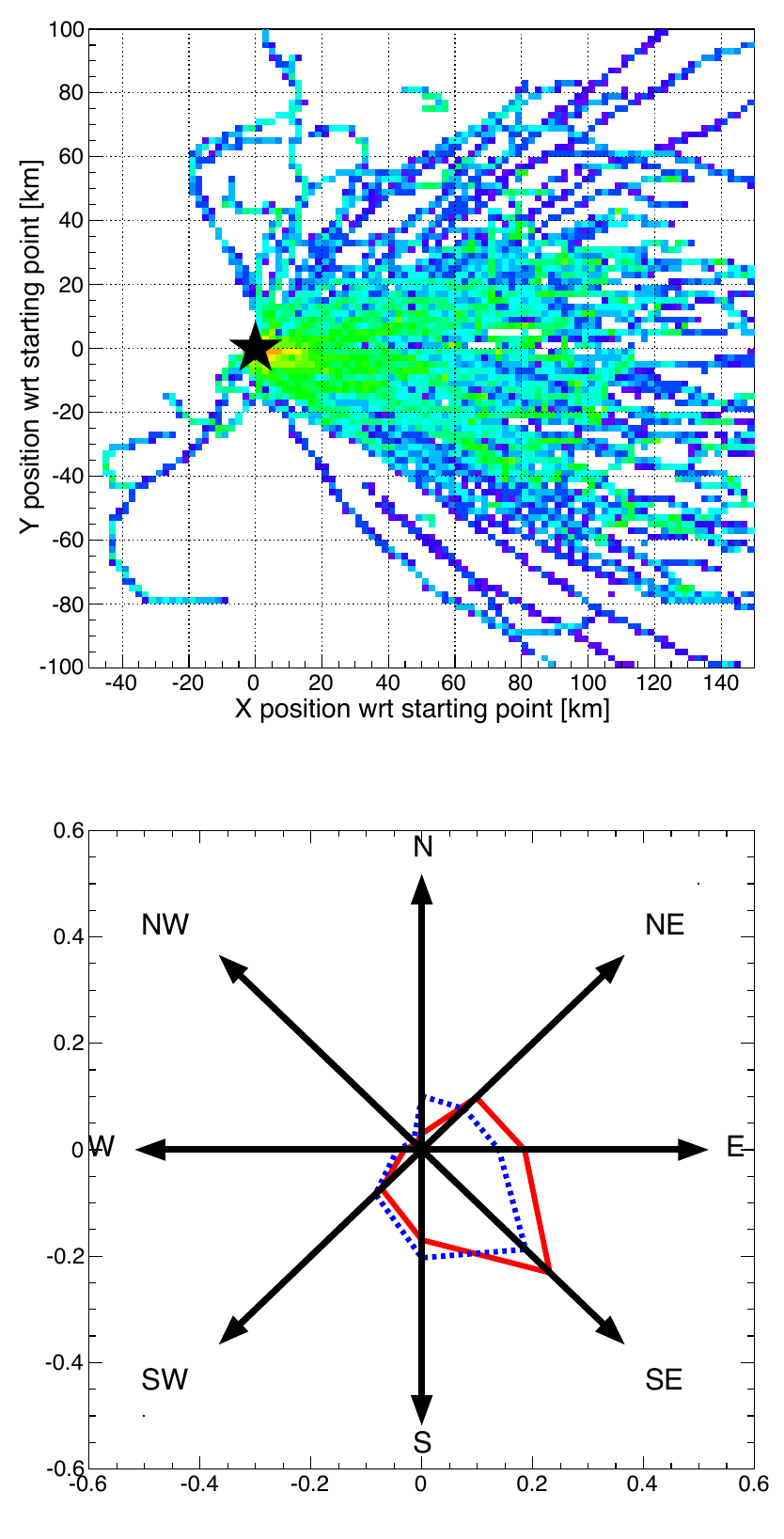}
\caption{{\bf Distribution of trajectories and direction of balloon flights at the Pierre Auger Observatory.} Balloon data set from August 2002 to December 2010. (top) Distribution of balloon trajectories with a start location fixed at $(0,0)$ and represented by a star. All the altitudes probed during the flight are integrated in this histogram. Colours indicate the frequency of a region, from red (more likely) to blue (less likely). (bottom) Directions of balloon flights are given in {\it dashed line}. Directions of air masses starting at 500~m above the observatory location ($35^\circ$ {{S}}, $69^\circ$ {{W}}) using HYSPLIT are plotted in {\it solid line} for the year 2008. Colours in online version.
}
\label{fig:HYSPLIT_vs_BALLOON}
\end{figure}

As mentioned in Section~\ref{sec:hysplit}, the HYSPLIT tool requires meteorological data from the GDAS model. Using the meteorological radio soundings performed at the Pierre Auger Observatory, a balloon track is available for each flight. In Figure~\ref{fig:Wind_profile_seasonal}, the average-vertical profiles of wind speed for each season are displayed, as measured during balloon flights at the observatory. Each of them is compared to the mean vertical profile extracted from GDAS data of the corresponding season. The wind speed fluctuates strongly day-by-day: the largest variations are measured in the Austral winter. In table~\ref{tab:stat_agree}, the mean values and the standard deviation values for the difference between measured radio sounding data and GDAS data for temperature, pressure, vapour pressure and wind speed are given. Concerning the wind speed which will be of primary interest in this work, we can see that its value is slightly underestimated by the GDAS model in the lower part of the atmosphere.

To validate the wind directions used in HYSPLIT calculations, the agreement between the directions of the balloon flights and the directions of air mass paths estimated using HYSPLIT is checked. In Figure~\ref{fig:HYSPLIT_vs_BALLOON}~(top), the distribution of balloon trajectories obtained at the Auger site is given. In this plot, the altitude evolution through the flight is not indicated. The corresponding directions of these balloon trajectories are given in blue in Figure~\ref{fig:HYSPLIT_vs_BALLOON}~(bottom), tending roughly to a South-East direction (detailed explanations on how to obtain this plot are given in Sect.~\ref{sec:methodology_aerosol}, where the steps here are given by the different data points recorded during the balloon flight). To exclude altitudes too much higher than the $500~$m AGL computed by HYSPLIT, only the first $20~$min of each flight is used to estimate the direction of a radio sounding. On the other hand, using the HYSPLIT tool, 48-h forward trajectories from an altitude fixed at $500~$m are computed every hour, for the year 2008. Following the same method as the one explained in Sect.~\ref{sec:methodology_aerosol}, the resulting distribution of air mass directions is plotted in red. The distributions for different initial altitudes will be shown later in Figure~\ref{fig:HYSPLIT_direction_fAltitude_2008}. Air mass directions are just slightly dependent on the altitude. The agreement between balloon trajectories and forward trajectories computed by HYSPLIT is once again very good: since a change in direction is not common for air mass trajectories at these probed altitudes, an agreement of directions along this short path can be extrapolated to larger distances travelled by air masses. Thus, after these two crosschecks, i.e.\ the vertical profiles for wind speed and the directions of air masses travelling above the Auger array, it can be concluded that air mass calculations for the location ($35^\circ$ {{S}}, $69^\circ$ {{W}}) are suitable. This conclusion complements the analysis and results of the former study of GDAS data for the Pierre Auger Observatory~\citep{GDASpaper}.

\subsection{Origin and trajectories of air masses arriving at the Pierre Auger Observatory}
\label{sec:methodology_aerosol}

\begin{table*}
\centering
\begin{tabular}{c c}
\toprule
Parameter & Setting \\[0.5ex]
\midrule
Meteorological dataset & GDAS\\
Trajectory direction & Backward / Forward\\
Trajectory duration & $48~$hours\\
Start point & Auger Observatory ($35^\circ$ {{S}}, $69^\circ$ {{W}})\\
Start height & $500~$m / $1000~$m AGL\\
Vertical motion & Model vertical velocity\\[1.0ex]
\bottomrule
\end{tabular}
~\\
\caption{{\bf Input parameters used for all HYSPLIT runs.}}
\label {tab:HYSPLIT_parameters}
\end{table*}

\begin{figure}[!t]
\centering
	\includegraphics [scale=0.46]{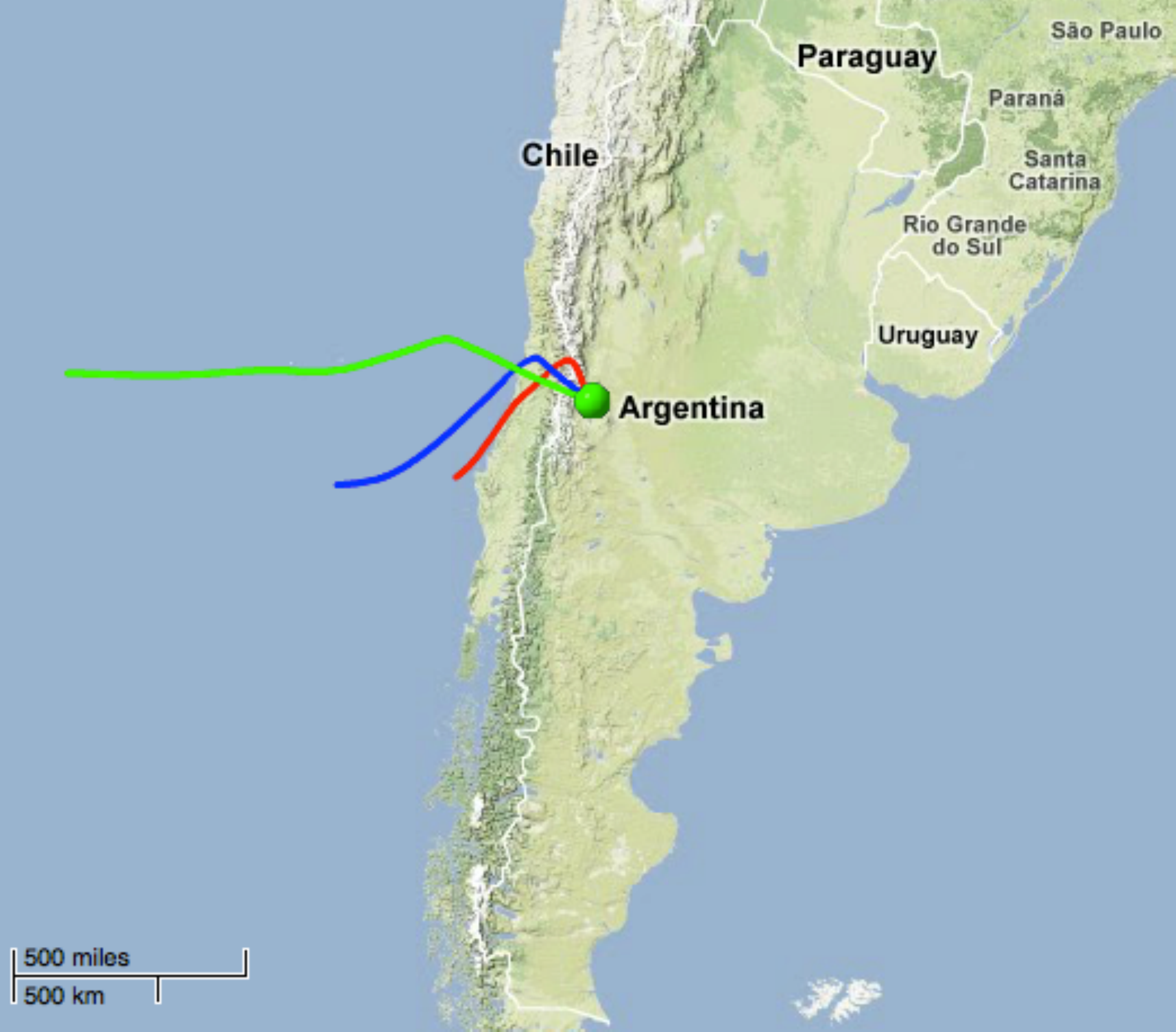}
\caption {{\bf Example of a back trajectory from the Malarg\"ue location using HYSPLIT.} The initial height is fixed at $500~$m above ground level for the {\it red} line, $1000~$m for the {\it blue} line and $3000~$m for the {\it green} line (map taken from Google Earth). Colours in online version.}
\label{fig:HYSPLIT_googlemap}
\end{figure}

The trajectories of air masses arriving at the Auger Observatory have been evaluated for eight years (2005--2012). In this way, the seasonal variations in the origin of air masses can be shown by the analysis. A 48-h backward trajectory is computed every hour, throughout the years. Also, the evolution profiles of the different meteorological quantities can be estimated and recorded. The key input parameters for the runs are given in Table~\ref{tab:HYSPLIT_parameters}. In Figure~\ref{fig:HYSPLIT_googlemap}, an example of a \mbox{$48$-h} backward trajectory from HYSPLIT is shown for altitudes fixed at $500~$m, $1000~$m or $3000~$m above the Malarg\"ue location. A 2-day time scale is a good compromise with respect to aerosol lifetime, air mass dispersion and computing time. Each run provides the geographical location of air mass trajectories (arriving at different altitudes) at the Auger Observatory and the evolution (along the trajectories) of the relevant meteorological physical parameters (temperature, relative humidity, etc). Some geographical locations of air masses show significant changes of direction during the previous $48$ hours. Two different methods have been used to display air mass origin and mean trajectory.

\begin{figure*}
\centering
	\includegraphics [scale=1.0]{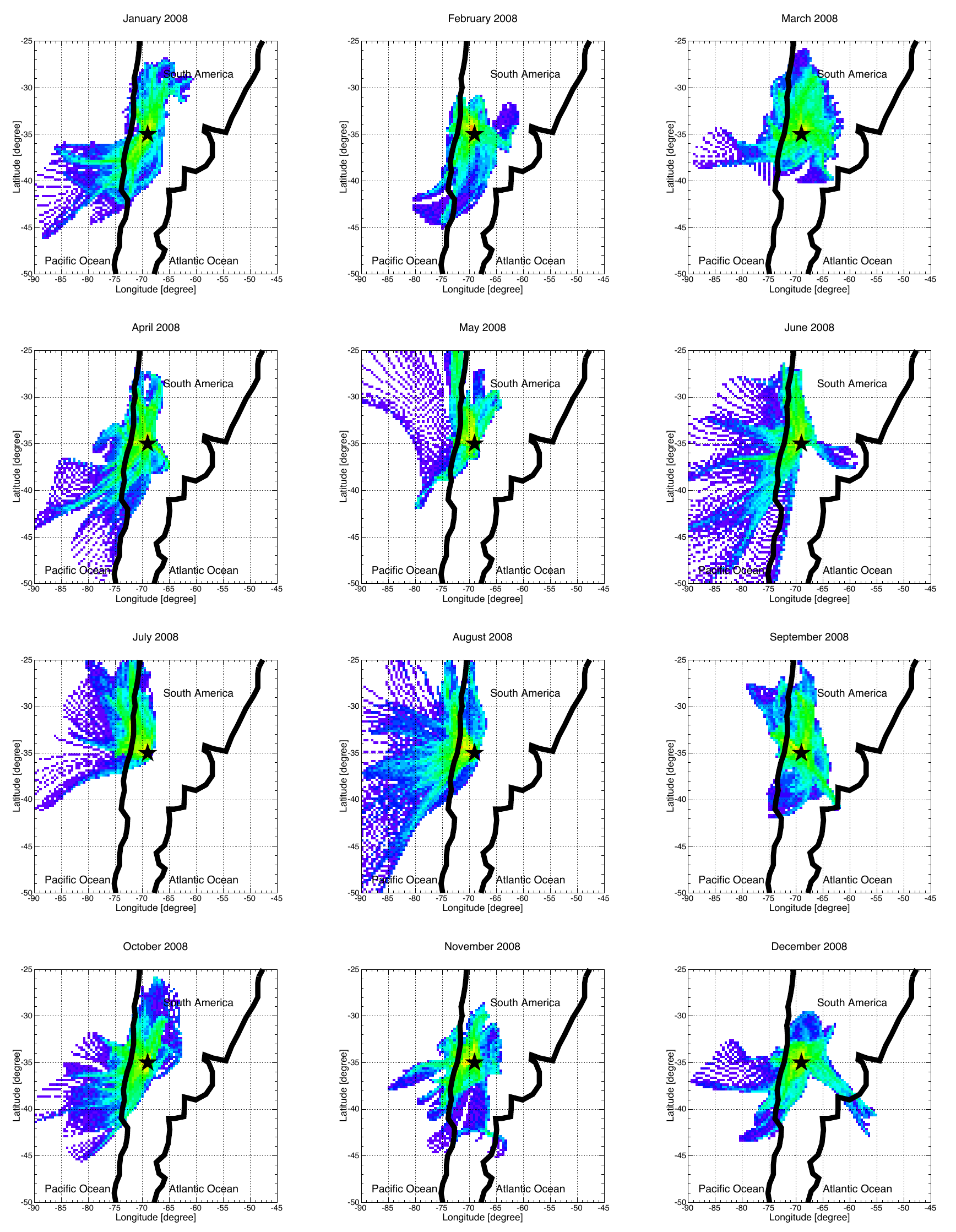}
\caption{{\bf Distribution of 48-h backward trajectories from the Malarg\"ue region by month.} Paths estimated with HYSPLIT for the year 2008, for a start altitude fixed at $500~$m AGL. The black star represents the Pierre Auger Observatory. The black line represents the South American coast. Colours indicate the frequency of a region, from red (more likely) to blue (less likely). Colours in online version.}
\label{fig:HYSPLIT_trajectories_2008}
\end{figure*}

The first visualisation is a two-dimensional diagram, longitude versus latitude. From this display, it is possible to extract regional influence on air quality at the Auger site. Also, changes in direction are obvious. Therefore all regions that an air mass path traversed during its entire $48$-h travel period towards the Pierre Auger Observatory are displayed. In Figure~\ref{fig:HYSPLIT_trajectories_2008}, the distribution of the backward trajectories for each month during the year 2008 is displayed, for a start altitude fixed at $500~$m above ground level at the observatory. Also the fluctuations change month-by-month: e.g., June or August are months where the air masses show large fluctuations trajectory-to-trajectory. These two months are exactly the ones having the highest fractions of clear profiles for this year in Table~\ref{tab:vaod_stats}. 

\begin{figure*}
\centering
	\includegraphics [scale=1.0]{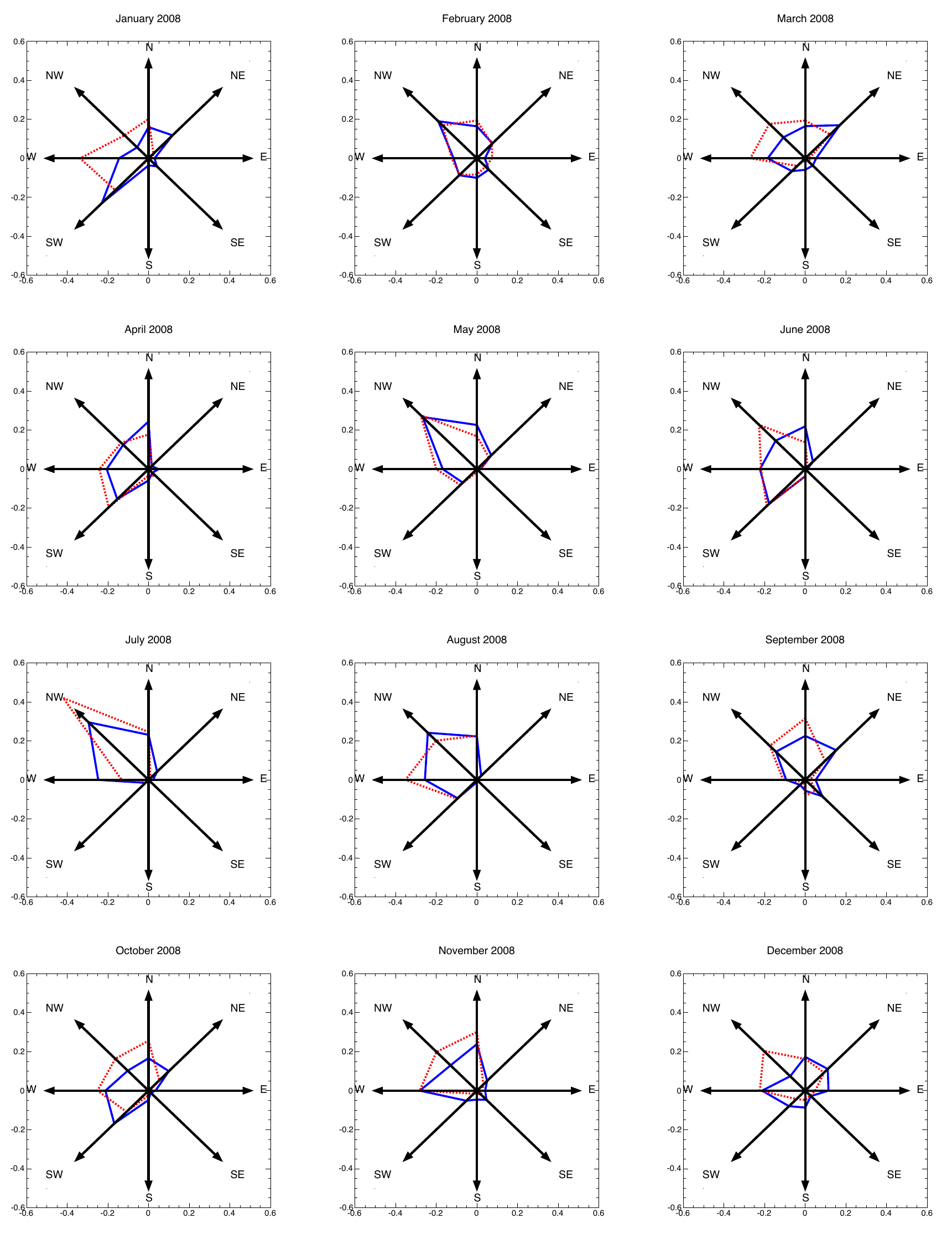}
\caption{{\bf Direction of air masses influencing the Auger atmosphere for each month.} Direction of trajectories estimated using HYSPLIT for the year 2008, with input parameters given in Table~\ref{tab:HYSPLIT_parameters}, at two different start altitudes: 500~m AGL ({\it solid line}) and 1000~m AGL ({\it dashed line}). Each distribution is normalised to one. Colours in online version.}
\label{fig:HYSPLIT_direction_fAltitude_2008}
\end{figure*}

Another visualisation of the trajectory is done using the direction of the air mass paths: it consists of subdividing each air mass trajectory into two 24-h sub-trajectories. The direction for the most recent sub-trajectory is then chosen among these directions: North/N ($0^\circ \pm 22.5^\circ$), North-East / NE ($45^\circ \pm 22.5^\circ$), East / E ($90^\circ \pm 22.5^\circ$), South-East / SE ($135^\circ \pm 22.5^\circ$), South / S ($180^\circ \pm 22.5^\circ$), South-West / SW ($225^\circ \pm 22.5^\circ$), West / W ($270^\circ \pm 22.5^\circ$), North-West / NW ($315^\circ \pm 22.5^\circ$) -- origin of the frame being fixed at the Pierre Auger Observatory. For each trajectory, its origin is obtained as follows: using the angle between two steps along the trajectory, a sub-direction is defined for each step (i.e.\ 24 in our case) and then the global direction corresponds to the most probable value of sub-directions for the whole sub-path. The different directions recorded are then plotted in a histogram. In Figure~\ref{fig:HYSPLIT_direction_fAltitude_2008}, these polar histograms are shown for each month of the year 2008. The polar histograms are normalised to one, i.e.\ the sum of the height entries corresponding to the height directions is equal to one. Most of the months have air masses with a North-West origin. Air masses coming from the East are particularly rare. The observations remain the same when the start altitude of the backward trajectories is modified. For the highest initial altitude ($1000~$m above ground level at the observatory), the fluctuations trajectory-to-trajectory are larger and the air masses travel faster; their endpoint is farther from the Pierre Auger Observatory.

\begin{figure*}[!t]
\centering
	\includegraphics [scale=1.0]{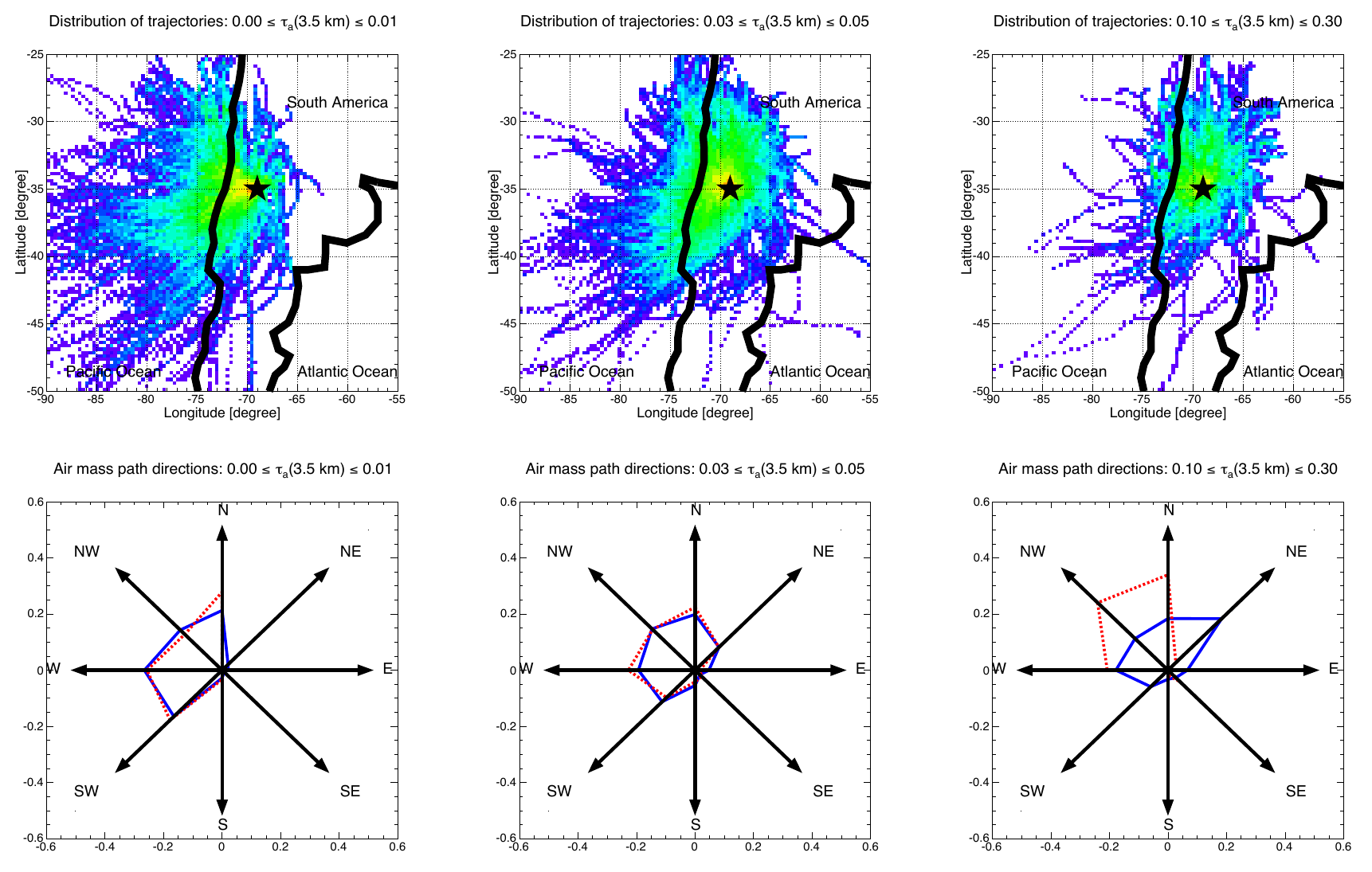}
\caption{{\bf Distribution of backward trajectories and direction of air masses for clear, average and hazy aerosol conditions.} Paths estimated with HYSPLIT for the years between 2005 and 2012 and aerosol optical depth data coming from the CLF measurements. (top) Distribution of 48-h backward trajectories from the Malarg\"ue location for a start altitude fixed at $500~$m AGL. Colours indicate the frequency of a region, from red (more likely) to blue (less likely). (bottom) Direction of trajectories with input parameters given in Table~\ref{tab:HYSPLIT_parameters}, at two different start altitudes AGL: 500~m ({\it solid line}) and 1000~m ({\it dashed line}). Each distribution is normalised to one. Colours in online version.
}
\label{fig:HYSPLIT_trajectories_VAOD}
\end{figure*}

\section{Interpretation of aerosol measurements using backward trajectories of air masses}
\label{sec:interpret_aerosol_hysplit}
As described in Section~\ref{sec:vaod_measurements}, aerosol concentrations measured at the Pierre Auger Observatory can fluctuate strongly night-by-night. Nevertheless, a seasonal trend with a minimum in Austral winter is found. The computed HYSPLIT trajectories are given in Figure~\ref{fig:HYSPLIT_trajectories_VAOD}~(top) for the conditions described in Section~\ref{sec:vaod_measurements} (clear, hazy and average hourly aerosol profiles). During clear conditions, the air masses come mainly from the Pacific Ocean as already observed in~\citet{AllenEtAl}. For hazy conditions, these air masses travel principally through continental areas during the previous 48~hours. Following the conclusion of a chemical aerosol analysis performed at the Auger site by \citet{MariaI_LongPaper}, NaCl crystals are detected in aerosol samplings during Austral winter, the period with mostly clear conditions and trajectories pointing back to the Pacific Ocean. Thus, these NaCl crystals could come from the Pacific Ocean, even if we cannot exclude another main origin as salt flats as main origin. Snow is another phenomenon that has to be taken into account here. As explained in~\citet{MariaI_LongPaper}, even though snowfalls are quite rare during winter in this region, the low temperatures conserve the snow on ground for long periods. An aerosol source (soil) blocked by snow, combined with air masses coming from Pacific Ocean, are probably the main causes for these clear nights observed in winter at the Pierre Auger Observatory.

The direction of air masses for clear, average and hazy aerosol conditions are presented in Figure~\ref{fig:HYSPLIT_trajectories_VAOD}~(bottom) for two different start altitudes at the location of the observatory: $500~$m AGL (solid line) and $1000~$m AGL (dashed line). The same conclusions can be drawn for both initial altitudes. Nevertheless, a slight shift to the West is seen for the highest altitude for hazy aerosol conditions. To relativise this disagreement, it seems important to note that atmospheric aerosols are usually located in the low part of the atmosphere, typically in the first 2~km. Hazy aerosol profiles are dominated by winds coming from the North-East (especially at low start altitude), typical in September/October, just after the Austral winter and its corresponding air masses coming from the Pacific Ocean. Such an observation of an increase in the aerosol optical depth values from August to September, combined with air masses coming from the North-East, could indicate a relationship to pollution emitted by surrounding urban areas (e.g. San Rafael or Mendoza) or the phenomenon of biomass burning in the northern part of Argentina or Bolivia. It is now well-known that wildfire emissions in South America, mainly in Brazil, Argentina, Bolivia or Paraguay, occurring during the dry season, strongly affect a vast part of the atmosphere in South America via long-term transportation of air masses~\citep{Andreae,Fiebig,Freitas,Ulke}. Locally, aerosol optical depth values are usually 10 to 20 times greater than in months without burning biomass. An analysis done by~\citet{CastroVidelaEtAl} shows that the mean number of fires reaches a peak from August to October in the northern part of South America ($0^\circ$ {S} to $30^\circ$ {S} in latitude). This fire activity could correspond to the peak observed in aerosol concentration at the Pierre Auger Observatory. However, without a more complex modelling of air masses, only hints can be given to explain this high fraction of hazy conditions in September and a more quantitative assessment of the link source/receptor would require further analysis.

\section{Conclusion}
A better understanding of air mass behaviour affecting the Pierre Auger Observatory has been realised by applying the HYSPLIT tool that tracks air masses from one region to another.  Air masses above the observatory do not have the same origin throughout the year. Aerosol concentrations measured at the observatory
depict two notable features:  a seasonal trend with a minimum reached in Austral winter, and a quick increase occurring yearly just after August. The first can be explained by air masses transported from the Pacific Ocean and travelling above snowy soils to the observatory. The peak in September and October could be interpreted as air mass transport from biomass burning occurring in the northern of South America (mainly in the northern of Argentina and Bolivia) during dry season. However, another cause such as air pollution transported from closer urban areas, also located to the north of the observatory, cannot be excluded. Future studies that include satellite data or ground-level monitoring between the observatory and possible pollution source regions could resolve this issue. However, for both cases, air mass transport plays a key role in the aerosol component present above the Pierre Auger Observatory.

\section{Acknowledgments}
The successful installation, commissioning, and operation of the Pierre Auger Observatory would not have been possible without the strong commitment and effort
from the technical and administrative staff in Malarg\"ue. G. Curci was supported by the Italian Space Agency (ASI) in the frame of PRIMES project.

\begin{sloppypar}
We are very grateful to the following agencies and organizations for financial support: 
Comisi\'on Nacional de Energ\'ia At\'omica, 
Fundaci\'on Antorchas,
Gobierno De La Provincia de Mendoza, 
Municipalidad de Malarg\"ue,
NDM Holdings and Valle Las Le\~nas, in gratitude for their continuing
cooperation over land access, Argentina; 
the Australian Research Council;
Conselho Nacional de Desenvolvimento Cient\'ifico e Tecnol\'ogico (CNPq),
Financiadora de Estudos e Projetos (FINEP),
Funda\c{c}\~ao de Amparo \`a Pesquisa do Estado de Rio de Janeiro (FAPERJ),
S\~ao Paulo Research Foundation (FAPESP) Grants \#2010/07359-6, \#1999/05404-3,
Minist\'erio de Ci\^{e}ncia e Tecnologia (MCT), Brazil;
AVCR, MSMT-CR LG13007, 7AMB12AR013, MSM0021620859, and TACR TA01010517 , Czech Republic;
Centre de Calcul IN2P3/CNRS, 
Centre National de la Recherche Scientifique (CNRS),
Conseil R\'egional Ile-de-France,
D\'epartement  Physique Nucl\'eaire et Corpusculaire (PNC-IN2P3/CNRS),
D\'epartement Sciences de l'Univers (SDU-INSU/CNRS), France;
Bundesministerium f\"ur Bildung und Forschung (BMBF),
Deutsche Forschungsgemeinschaft (DFG),
Finanzministerium Baden-W\"urttemberg,
Helmholtz-Gemeinschaft Deutscher Forschungszentren (HGF),
Ministerium f\"ur Wissenschaft und Forschung, Nordrhein-Westfalen,
Ministerium f\"ur Wissenschaft, Forschung und Kunst, Baden-W\"urttemberg, Germany; 
Istituto Nazionale di Fisica Nucleare (INFN),
Ministero dell'Istruzione, dell'Universit\`a e della Ricerca (MIUR), 
Gran Sasso Center for Astroparticle Physics (CFA), CETEMPS Center of Excellence, Italy;
Consejo Nacional de Ciencia y Tecnolog\'ia (CONACYT), Mexico;
Ministerie van Onderwijs, Cultuur en Wetenschap,
Nederlandse Organisatie voor Wetenschappelijk Onderzoek (NWO),
Stichting voor Fundamenteel Onderzoek der Materie (FOM), Netherlands;
Ministry of Science and Higher Education,
Grant Nos. N N202 200239 and N N202 207238, 
The National Centre for Research and Development Grant No ERA-NET-ASPERA/02/11, Poland;
Portuguese national funds and FEDER funds within COMPETE - Programa Operacional Factores de Competitividade through 
Funda\c{c}\~ao para a Ci\^{e}ncia e a Tecnologia, Portugal;
Romanian Authority for Scientific Research ANCS, 
CNDI-UEFISCDI partnership projects nr.20/2012 and nr.194/2012, 
project nr.1/ASPERA2/2012 ERA-NET, PN-II-RU-PD-2011-3-0145-17, and PN-II-RU-PD-2011-3-0062, Romania; 
Ministry for Higher Education, Science, and Technology,
Slovenian Research Agency, Slovenia;
Comunidad de Madrid, 
FEDER funds, 
Ministerio de Ciencia e Innovaci\'on and Consolider-Ingenio 2010 (CPAN),
Xunta de Galicia, Spain;
The Leverhulme Foundation,
Science and Technology Facilities Council, United Kingdom;
Department of Energy, Contract Nos. DE-AC02-07CH11359, DE-FR02-04ER41300, DE-FG02-99ER41107,
National Science Foundation, Grant No. 0450696, 0855680, 1207605,
The Grainger Foundation USA; 
NAFOSTED, Vietnam;
Marie Curie-IRSES/EPLANET, European Particle Physics Latin American Network, 
European Union 7th Framework Program, Grant No. PIRSES-2009-GA-246806; 
and UNESCO.
\end{sloppypar}

\appendix

\section{Supplementary data}
Supplementary data to this article can be found online.

\bibliographystyle{elsarticle-harv}



\clearpage
\begin{sloppypar}
\par\noindent
{\bf The Pierre Auger Collaboration} \\
A.~Aab$^{42}$, 
P.~Abreu$^{65}$, 
M.~Aglietta$^{54}$, 
M.~Ahlers$^{95}$, 
E.J.~Ahn$^{83}$, 
I.~Al Samarai$^{29}$, 
I.F.M.~Albuquerque$^{17}$, 
I.~Allekotte$^{1}$, 
J.~Allen$^{87}$, 
P.~Allison$^{89}$
% ...

\par\noindent
% ...
$^{1}$ Centro At\'{o}mico Bariloche and Instituto Balseiro (CNEA-UNCuyo-CONICET), San 
Carlos de Bariloche, 
Argentina \\
% ...
$^{17}$ Universidade de S\~{a}o Paulo, Instituto de F\'{\i}sica, S\~{a}o Paulo, SP, 
Brazil \\
% ...
$^{29}$ Institut de Physique Nucl\'{e}aire d'Orsay (IPNO), Universit\'{e} Paris 11, CNRS-IN2P3, 
Orsay, 
France \\
% ...
$^{42}$ Universit\"{a}t Siegen, Siegen, 
Germany \\
% ...
$^{54}$ Osservatorio Astrofisico di Torino  (INAF), Universit\`{a} di Torino and Sezione INFN, 
Torino, 
Italy \\
% ...
$^{65}$ Laborat\'{o}rio de Instrumenta\c{c}\~{a}o e F\'{\i}sica Experimental de Part\'{\i}culas - LIP and  
Instituto Superior T\'{e}cnico - IST, Universidade de Lisboa - UL, 
Portugal \\
% ...
$^{83}$ Fermilab, Batavia, IL, 
USA \\
% ...
$^{87}$ New York University, New York, NY, 
USA \\
% ...
$^{89}$ Ohio State University, Columbus, OH, 
USA \\
% ...
$^{95}$ University of Wisconsin, Madison, WI, 
USA \\
% ...
\par\noindent
(\ddag) Deceased \\
(a) Now at Konan University \\
(b) Also at the Universidad Autonoma de Chiapas on leave of absence from Cinvestav \\
(c) Now at NYU Abu Dhabi \\
\end{sloppypar}

\end{document}